# Tuning perpendicular magnetic anisotropy in (Ga,Mn)(As,P) by thermal annealing

A. Casiraghi, A.W. Rushforth, M. Wang, N. R. S Farley, P. Wadley, J. L. Hall, C.R. Staddon, K.W. Edmonds, R.P. Campion, C.T. Foxon, B.L. Gallagher.

School of Physics and Astronomy, University of Nottingham, University Park, Nottingham, NG7 2RD United Kingdom

*We have investigated the effects of post growth low temperature annealing on the magnetic, electrical and structural properties of $(Ga_{0.94},Mn_{0.06})(As_{0.9},P_{0.1})$ layers grown by molecular beam epitaxy. By controlling the annealing time we are able to tune the magnetic anisotropy between an easy axis in the plane for the as-grown samples, to an easy axis perpendicular to the plane for fully annealed samples. The increase of the carrier density, as a result of annealing, is found to be the primary reason for the change in magnetic anisotropy, in qualitative agreement with theoretical predictions.*

The III-V ferromagnetic semiconductor (Ga,Mn)As, alloyed with phosphorus, has been the subject of recent investigations of its magnetic and electrical transport properties[1-5]. In addition to the favorable properties of (Ga,Mn)As and its alloys, such as a high sensitivity of the magnetic properties to epitaxial strain and the Mn and carrier concentrations[6,7], the (Ga,Mn)(As,P) compound has proved to be of particular interest because it is possible to produce layers with a perpendicular to plane magnetic easy axis by selecting the P concentration such that the layers are under tensile strain when grown on a GaAs substrate. This property is beneficial for many experiments because it allows magnetization switching and domain wall motion to be detected using Kerr microscopy[8,9] and the anomalous Hall effect[8,10]. Also, these layers do not suffer from the problem of pinning sites formed by dislocations that arise in other layer structures designed to produce perpendicular to plane anisotropy, such as (Ga,Mn)As grown on a relaxed InGaAs buffer layer[11].

Previously, we investigated the properties of $(Ga_{1-x},Mn_x)(As_{1-y},P_y)$ layers grown by molecular beam epitaxy (MBE) for $y \geq 0.1$[1] and showed how the perpendicular to plane anisotropy became stronger as the tensile strain was increased by increasing y. Other studies investigated the properties of the compound for $y<0.1$[2,3] and showed that the anisotropy switches from in plane easy axis to perpendicular to plane easy axis as the growth strain is varied from compressive to tensile, with the boundary at $y \approx 0.05$. These previous studies focused on samples that had been subjected to a post growth low temperature annealing step to remove interstitial Mn ions ($Mn_I$) from the layer. $Mn_I$ are known to act as compensating defects in (Ga,Mn)As alloys resulting in a decrease of the hole density. They also couple antiferromagnetically with the substitutional Mn ions. Therefore, as-grown samples have a lower moment density and a lower Curie temperature than annealed samples[12]. The magnetic anisotropy is also expected to depend sensitively on the carrier density in (Ga,Mn)As, with the easy axis rotating from in plane to out of plane as a function of increasing carrier density [6] for layers grown under tensile strain. In this paper we investigate the effects of low temperature annealing on the magnetic and transport properties of a $(Ga_{0.94}Mn_{0.06})(As_{0.9}P_{0.1})$ layer. We show that, by varying the time of annealing, it is possible to tune the magnetic anisotropy of the layer so that the magnetic easy axis can be rotated from in plane for the as-grown sample to out of plane for the annealed sample.

The sample was grown by molecular beam epitaxy on a [001] GaAs substrate. A 50nm low temperature $GaAs_{0.9}P_{0.1}$ buffer layer was grown immediately before a 25nm $Ga_{0.94}Mn_{0.06}As_{0.9}P_{0.1}$ layer. Further details of the growth procedure are published elsewhere [1]. The magnetic properties were measured using a Quantum Design MPMS SQUID magnetometer. X-Ray Diffraction (XRD) measurements were obtained using a Philips X'Pert Materials Research diffractometer. Four terminal electrical transport measurements were carried out on Hall bars fabricated using photolithography.

Figures 1 (a) and (b) show magnetic hysteresis loops measured at 2K for the as-grown and fully annealed samples respectively. For the annealed sample the magnetic easy axis is perpendicular to the plane of the layer, as reported previously[1] and evidenced by the square M v H loop for the field applied in this direction, and by the sloping M v H curve and negligible remanance for the field applied in the plane of the layer. In contrast, for the as-grown sample the easy axis is predominantly in the plane, as evidenced by the square M v H loop for the field applied in the in-plane [110] direction. For the field applied in the out of plane direction, the small remanance indicates that this is a hard axis direction. The open loop sections of the M v H curve at higher fields indicate that the magnetic anisotropy energy has a local stable minimum in the out of plane direction when a finite field is applied. The magnetic reversal then proceeds via two steps, the first involving a rotation from out of plane to in plane as the field is reduced, followed by a rotation to out of plane as the field is increased in the reverse direction. A similar multistep reversal mechanism has been observed previously by Liu et al [13] for (Ga,Mn)As with low Mn concentration. It is interesting to note that for the as-grown sample the in-plane magnetic anisotropy consists of a superposition of a cubic [100]/[010] anisotropy term and a uniaxial anisotropy term favoring an easy axis along the [110] direction. This is similar to the magnetic anisotropy often observed in (Ga,Mn)As samples, except that the uniaxial anisotropy term usually favors a $[1\bar{1}0]$ direction in that case. We have studied the temperature dependence of the remnance (not shown) for $(Ga_{0.94},Mn_{0.06})(As_{1-y},P_y)$ samples with y in the range 0.03 to 0.3 and found that all as-grown samples show a combination of cubic [100]/[010] anisotropy and uniaxial easy [110] anisotropy at the lowest temperatures with the ratio of the uniaxial to cubic anisotropy increasing as y increases. For samples with y<0.1 the uniaxial anisotropy rotates from a [110] easy direction to a $[1\bar{1}0]$ easy direction as the temperature is increased. For y=0.03 and 0.06 we find that fully annealing the samples causes the uniaxial term to switch to a $[1\bar{1}0]$ easy direction over the whole temperature range below the Curie temperature. For these samples the magnetic anisotropy remains in the plane upon annealing. For the remainder of this manuscript we shall concentrate on the effects of annealing time on the sample with y=0.1, which shows a transition from in-plane to out of plane magnetic anisotropy upon annealing.

The XRD line scans in Fig. 2 indicate that the $(Ga_{0.94}Mn_{0.06})(As_{0.9}P_{0.1})$ layers are under tensile strain in both the as-grown and annealed samples. This is revealed by the broad peak from the $(Ga_{0.94}Mn_{0.06})(As_{0.9}P_{0.1})$ layer situated to the right hand side of the sharp GaAs substrate peak. The tensile strain is the origin of the perpendicular magnetic anisotropy in the annealed samples [1-3]. The small variation in the position of the $(Ga_{0.94}Mn_{0.06})(As_{0.9}P_{0.1})$ peak on annealing shows that the annealing procedure increases the tensile strain slightly, consistent with the removal of $Mn_I$ from the lattice.

The effect of the annealing time on the magnetic and transport properties was investigated using a Hall bar device fabricated from the same $(Ga_{0.94}Mn_{0.06})(As_{0.9}P_{0.1})$ wafer. Four terminal longitudinal and Hall resistances were measured in a perpendicular magnetic field after several successive annealing steps of different time periods at 180°C. The results are shown in Fig. 3. Annealing results in a decrease of the longitudinal resistivity and a gradual transition to a more metallic-like temperature dependence. This is accompanied by an increase of the Curie temperature, as deduced from the position of the peak in dR/dT [14] from each curve, summarised in the inset to Fig. 3(a). These trends are consistent with our understanding that the annealing causes the $Mn_I$ to be removed from the $(Ga_{0.94}Mn_{0.06})(As_{0.9}P_{0.1})$ layer by diffusion and passivation at the surface, giving rise to an increase in the carrier density. The measured Hall resistance is proportional to the perpendicular component of the magnetization, through the anomalous Hall effect [15]. Figure 3(b) shows the Hall resistance measured in perpendicular magnetic field at 1.5K. In the as-grown state the high resistivity prevents a proper measurement of the true Hall resistance. However, after only 20 minutes of annealing the decrease in the longitudinal resistivity allows the Hall resistance to be measured, and this closely resembles the M v H loop measured by SQUID magnetometry on the as-grown sample. Subsequent annealing steps result in a gradual transition from the sloping M v H loop to a square loop, reminiscent of that observed by SQUID magnetometry on the sample annealed for 48 hours. Indeed, after only 140 minutes the hysteresis loop is almost fully square, indicating that the film now possesses an easy axis perpendicular to the plane. Further annealing results in a small decrease of the coercive field, perhaps due to a dependence of the domain wall energy on the carrier density, or the removal of $Mn_I$ which may otherwise act as domain wall pinning sites. These observations show that by controlling the anneal time the magnetic anisotropy of the $(Ga_{0.94}Mn_{0.06})(As_{0.9}P_{0.1})$ layer can be tuned between an in plane easy axis and an out of plane easy axis. This is accompanied by a decrease in the longitudinal resistivity and an increase in the Curie temperature. Given that the layer is under tensile strain before and after annealing this suggests that the change in the anisotropy arises mainly as a result of an increase in the carrier density on annealing. Such behaviour is qualitatively consistent with the predictions of Dietl et al. [6] for the dependence of the magnetic anisotropy on carrier density for (Ga,Mn)As compounds under tensile strain.

In summary we have investigated the effects of post growth low temperature annealing on the magnetic, electrical and structural properties of $(Ga_{0.94},Mn_{0.06})(As_{0.9},P_{0.1})$ layers. The results suggest that the evolution of the magnetic anisotropy between an easy axis in the plane for the as-grown samples, to an easy axis perpendicular to the plane for fully annealed samples arises from the increase of the carrier density, as a result of annealing, This is in qualitative agreement with theoretical predictions of Dietl et al. [6].

We acknowledge funding from EU grants 214499-NAMASTE and 215368-SemiSpinNet, and EPSRC grants GR/S81407/01 and EP/D051487.

**Figure 1 (Color online)** Magnetic hysteresis loops measured by SQUID at T = 2 K with the external magnetic field applied in the film plane along the [110] direction (closed squares) and perpendicular to the plane (open circles) for (a) as-grown and (b) annealed $(Ga_{0.94},Mn_{0.06})(As_{0.9},P_{0.1})$ layers.

**Figure 2 (Color online)** XRD $\omega$-$2\theta$ scans for the $(Ga_{0.94},Mn_{0.06})(As_{0.9},P_{0.1})$ film before and after annealing for 48 hours. The two curves are offset for clarity. The gray curves show the best fit using Philips Expert Smoothfit software. The fitting reveals that the layers are under tensile strain with a lattice mismatch of 0.28% for the as grown sample and 0.49% after annealing.

**Figure 3 (Color online)** (a) Resistivity vs temperature for $(Ga_{0.94},Mn_{0.06})(As_{0.9},P_{0.1})$ and (b) anomalous Hall measurements (normalized to $R_{Hall}$ at $\mu_0 H=0.5T$) at T = 1.5 K for $(Ga_{0.94},Mn_{0.06})(As_{0.9},P_{0.1})$ as-grown and after successive annealing steps at 180°C. The external magnetic field is applied perpendicular to the film plane. The inset in (a) shows $T_C$ as a function of annealing time.

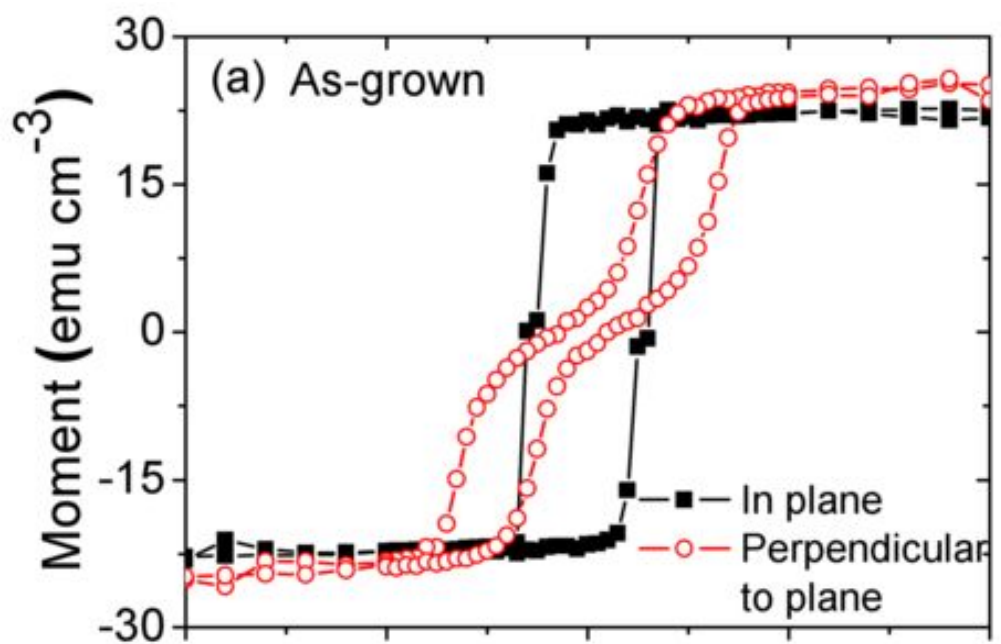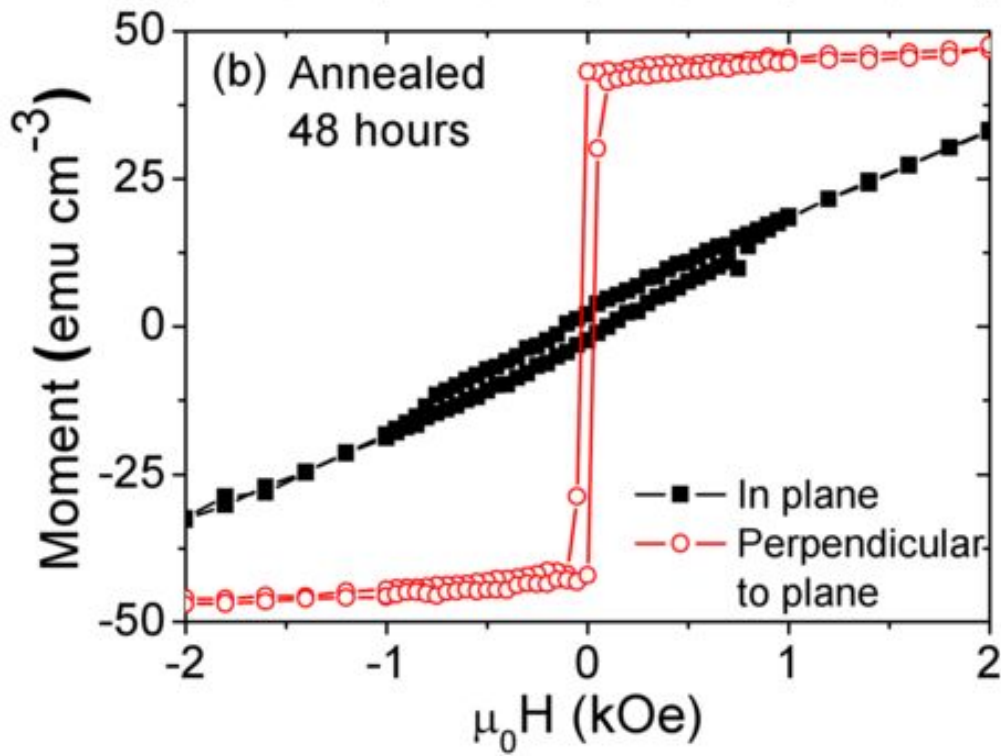

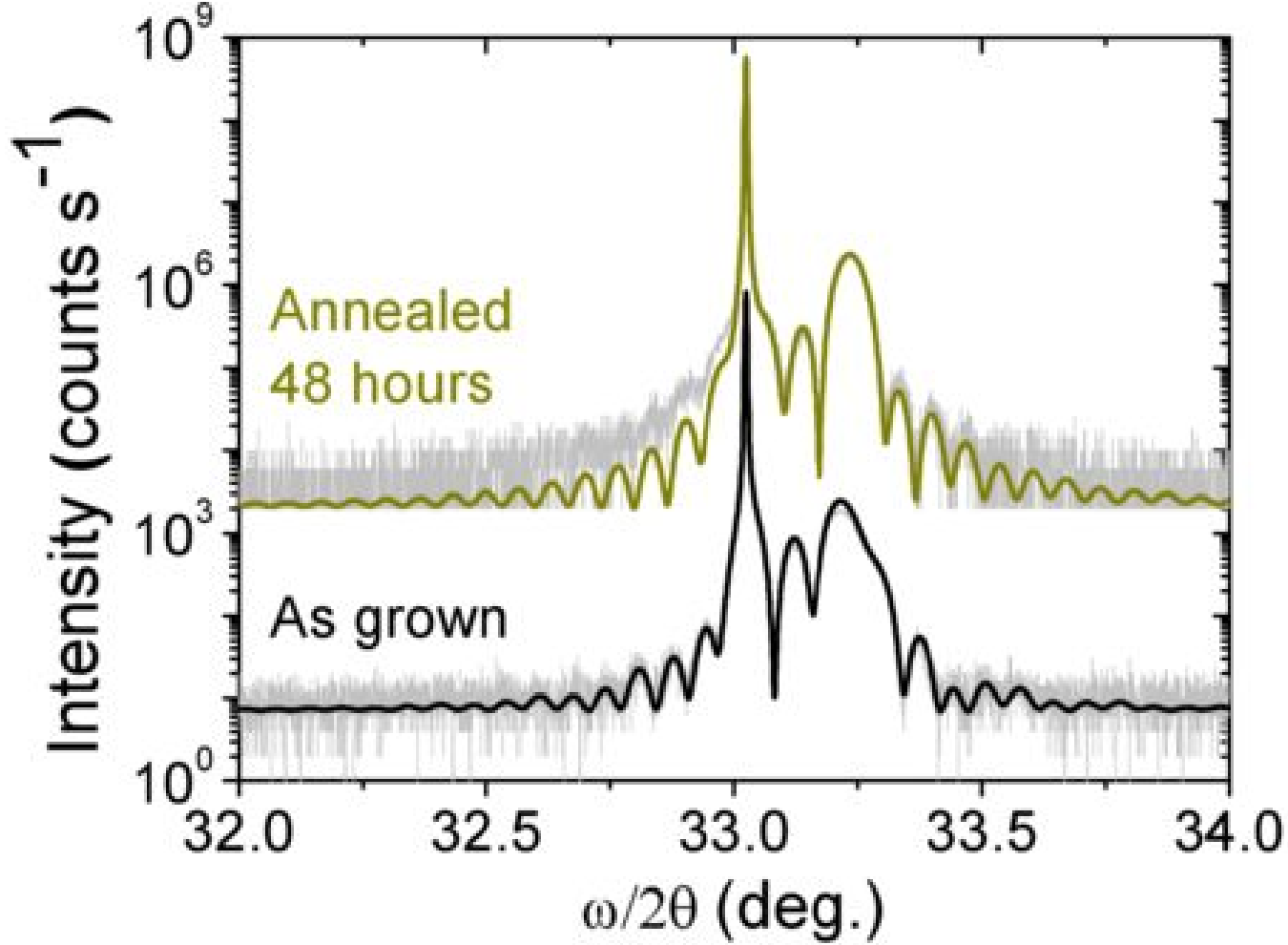

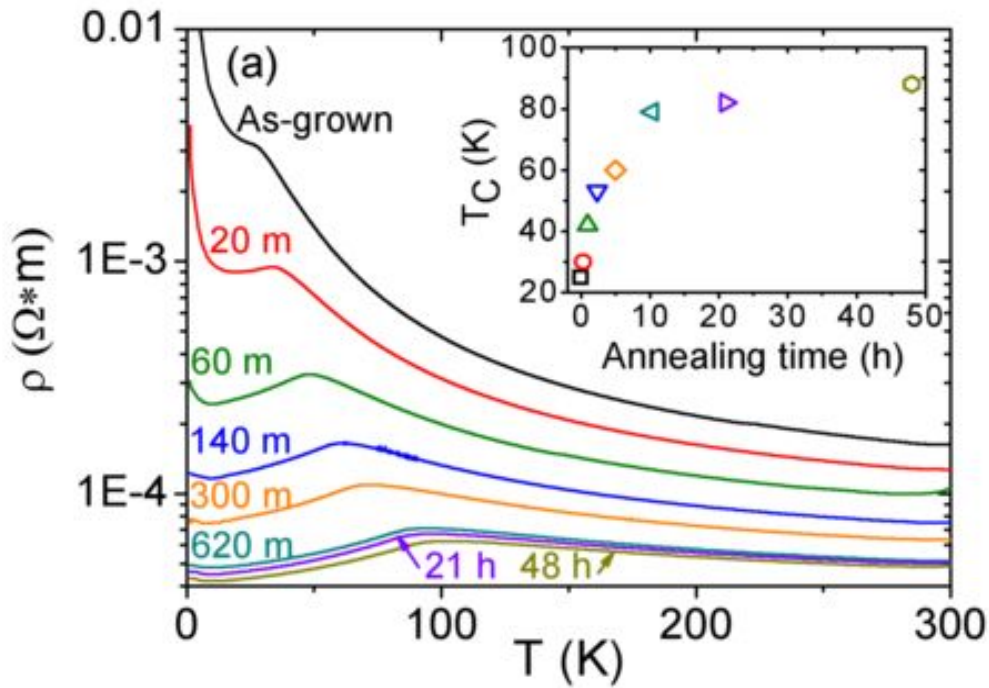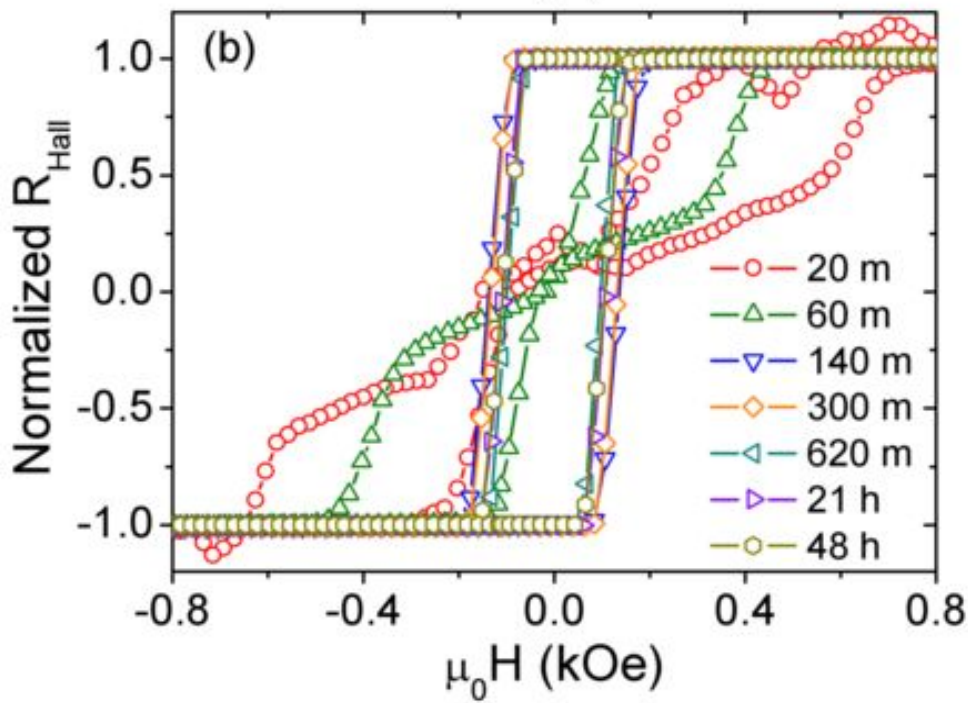